\documentclass[aps,pra,reprint,superscriptaddress,showpacs,longbibliography,floatfix,footnoteinbib]{revtex4-1}
\usepackage{amsmath,amssymb,graphicx,units}
\usepackage[plainpages=false,pdfpagelabels,colorlinks=true,linkcolor=red,
urlcolor=blue,citecolor=blue,pdftitle={Title},pdfauthor={},pdfdisplaydoctitle=true,
pdfduplex=DuplexFlipLongEdge]{hyperref}

\newcommand{\ket}[1]{{\vert #1\rangle}}

\newcommand{\1}{\mbox{\bf 1}}

\makeatother
\begin{document}

\title{Phase diagram of the interacting Haldane model with spin-dependent sublattice potentials}

\author{Can Shao}
\email{shaocan@njust.edu.cn}
\affiliation{Department of Applied Physics and MIIT Key Laboratory of Semiconductor Microstructure and Quantum Sensing, Nanjing University of Science and Technology, Nanjing 210094, China}


\author{Hong-Gang Luo}
\email{luohg@lzu.edu.cn}
\affiliation{School of Physical Science and Technology, Lanzhou University, Lanzhou 730000, China}
\affiliation{Lanzhou Center for Theoretical Physics $\&$ Key Laboratory of Theoretical Physics of Gansu Province, Lanzhou University, Lanzhou 730000, China}

\date{\today}

\begin{abstract}
Using the exact-diagonalization (ED) and mean-field (MF) approaches, we investigate the ground-state phase diagram of the interacting Haldane model on the honeycomb lattice, incorporating spin-dependent sublattice potentials $\Delta_{\sigma,\alpha}$. Here $\alpha=\text{A}$,$\text{B}$ and $\sigma=\uparrow$,$\downarrow$ denote the sublattice and spin components, respectively. Setting $\Delta_{\sigma,\text{A}}=+\Delta$ ($-\Delta$) and $\Delta_{\sigma,\text{B}}$$=-\Delta$ ($+\Delta$) for $\sigma=\uparrow$ ($\downarrow$) results in the system favoring a spin ordered state. Conversely, introducing the nearest-neighbor Coulomb interaction can induce charge ordering in the system. Due to the competition between these factors, we observe that in both ED and MF approaches, an exotic state with Chern number $C=1$ survives amidst two locally ordered phases and a topologically ordered phase with $C=2$. In the ED method, various  properties, such as the fidelity metric, the excitation gaps and the structure factors, are employed to identify critical points. In the MF method, using a sufficiently large lattice size, we define the local order parameters and band gaps to characterize the phase transitions. The interacting Haldane model and the spin-dependent lattice potential may be experimentally realized in an ultracold atom gas, providing a potential means to detect this intriguing state.
\end{abstract}

\maketitle

\section{Introduction}\label{sec:intro}

In contrast to the traditional framework of the Landau-Ginzburg theory, which relies on locally defined order parameters resulting from broken symmetries, topological phases have been identified and characterized based on their global, nonlocal properties\cite{Hasan_10,Qi_11}. Over the past few years, the categorization of topologically ordered states in non-interacting systems has been completed, considering various symmetries\cite{Chiu_16,Zhang19,Kruthoff2017,Vergniory19,Tang19,Schnyder08}.

On the other hand, there has been extensive research on the interacting topological insulators, which involve the interplay between topological properties and electronic correlations~\cite{Rachel_2018, Hohenadler_2013}. Correlation effects are expected to give rise to exotic states in the presence of topologically nontrivial conditions. Examples include the antiferromagnetic topological state in the  Bernevig-Hughes-Zhang model with the on-site Hubbard interaction\cite{Miyakoshi13, Yoshida13} and the topologically non-trivial phase with $C=1$ ($C$ denoting the Chern number) in the spinful Haldane-Hubbard model on honeycomb lattice\cite{He_2011,Zhu_2014,Wu_2015, Vanhala2016,Tupitsyn19,Mertz19,shao23, He_2024}. The origin of this $C=1$ phase is attributed to a spontaneous SU(2) symmetry breaking, with one spin component in the Hall state and the other in a localized state. Similar investigations include the antiferromagnetic Chern insulator in Kane-Mele-Hubbard model~\cite{Jiang18} and $C=1$ phase in interacting topological models on the square lattice~\cite{Wang2019, Ebrahimkhas21}. Additionally, introducing disorder to the interacting Haldane model~\cite{silva2023} or double exchange processes to the Haldane Hamiltonian~\cite{Tran_2022} can also break the spin symmetry. However, interplays between topology, on-site, and nearest-neighbor interactions do not exhibit such phenomenon~\cite{Wang2019, Shao2021}.
In experimental settings, observing these states remains challenging in realistic materials. Trapped cold atoms may offer an alternative and promising approach to achieving this purpose. Notably, the experimental realization of the topological Haldane model has been reported\cite{Jotzu2014}. Quantum simulations of strongly correlated systems in ultracold Fermi gases, including the Fermi-Hubbard model, have also been reviewed in Ref.~\cite{Hofstetter_2018,Esslinger2010,TARRUELL2018}.

In this paper, drawing inspiration from studies on spin-dependent optical lattice\cite{Jotzu15,Forster09,Mandel03}, we propose an interacting Haldane model on honeycomb lattice with spin-dependent sublattice potentials $\Delta_{\sigma,\alpha}$, where $\sigma=\uparrow,\downarrow$ and $\alpha=$A,B represent the spin and sublattice indices, respectively. Setting $\Delta_{\sigma,\text{A}}=+\Delta$ ($-\Delta$) and $\Delta_{\sigma,\text{B}}$$=-\Delta$ ($+\Delta$) for $\sigma=\uparrow$ ($\downarrow$) leads to the system favoring a staggered spin order (SSO) state. While introducing the nearest-neighbor Coulomb interaction $V$ drives the system into a staggered charge order (SCO) state. Due to their competitions, an intermediate state is expected and we explore the ground-state phase diagram of this model. Our findings reveal that, in addition to the topologically nontrivial phase with Chern number $C=2$ and two topologically trivial phases (SSO and SCO) with $C=0$, an newly generated phase with $C=1$ can be observed in both exact-diagonalization (ED) and mean-field (MF) methods. Various  properties, including the excitation gaps, the structure factors, and the fidelity metric in ED, as well as the band gap and local order parameters in MF, are obtained to characterize the phase transitions.
Especially, an effective potential differences for the two spin species in MF method are defined to analyze the origin of the $C=1$ phase.
Our finding presents a perspective on the interplay between topology, electronic correlation and lattice potential, shedding light on the realization of exotic states.

The presentation is structured as follows: we introduce the model, methods and relevant quantities in Sec.~\ref{sec:model}. Sections \ref{sec:results} presents our results based on the ED and MF approaches. Lastly, a conclusion is provided in Sec.~\ref{sec:conclusion}.

\section{Model and measurements} \label{sec:model}
The Hamiltonian of the interacting Haldane model with spin-dependent lattice potentials can be writen as
\begin{eqnarray}
\hat H=\hat H_k+\hat H_l,
\label{eq:H}
\end{eqnarray}
where the kinetic part
\begin{eqnarray}
\hat H_k=&-&t_1\sum_{\langle i,j\rangle,\sigma}(\hat c^{\dagger}_{i,\sigma} \hat c^{\phantom{}}_{j,\sigma}+\text{H.c.}) \nonumber \\
&-& t_2\sum_{\langle\langle i,j\rangle\rangle,\sigma}(e^{{\rm i}\phi_{ij}}\hat c^{\dagger}_{i,\sigma} \hat c^{\phantom{}}_{j,\sigma}+\text{H.c.})
\label{eq:H1}
\end{eqnarray}
and the local part
\begin{eqnarray}
\hat H_l&=&U\sum_{i}(\hat n_{i,\uparrow}-\frac{1}{2})(\hat n_{i,\downarrow}-\frac{1}{2})\nonumber \\
&+&V\sum_{\langle i,j\rangle}(\hat n_{i}-1)(\hat n_{j}-1)
+\sum_{i,\sigma}\Delta_{\sigma,\alpha}\hat n_{i,\sigma}.
\label{eq:H2}
\end{eqnarray}
In Eq.~(\ref{eq:H1}), $\hat c^{\dagger}_{i,\sigma}$ ($\hat c^{\phantom{}}_{i,\sigma}$) is the creation (annihilation) operator for an electron at site $i$ with spin $\sigma=\uparrow {\rm or} \downarrow$. $t_1$ ($t_2$) is the nearest-neighbor (next-nearest-neighbor) hopping constant and the Haldane phase $\phi_{i,j}=\phi$ ($-\phi$) in the clockwise (anticlockwise) loop is introduced to the next-nearest-neighbor hopping terms. In Eq.~(\ref{eq:H2}), $\hat n_{i,\sigma}= \hat c_{i,\sigma}^\dagger\hat c_{i,\sigma}^{\phantom{}}$ and $\hat n_{i}=\hat n_{i,\uparrow}+\hat n_{i,\downarrow}$; $U$ and $V$ are the on-site and nearest-neighbor Coulomb interactions, respectively.  $\Delta_{\sigma,\alpha}$ is the spin-dependent lattice potential: for spin $\sigma=\uparrow$, $\Delta_{\sigma,A}=+\Delta$ and $\Delta_{\sigma,B}=-\Delta$; for spin $\sigma=\downarrow$, $\Delta_{\sigma,A}=-\Delta$ and $\Delta_{\sigma,B}=+\Delta$.

In what follows, the model in Eq.~(\ref{eq:H}) is named as the extended Haldane-Hubbard model with spin-dependent lattice potentials. Throughout the paper, we set $t_1=1$, $t_2=0.2$, $\phi=\pi/2$ and focus on the ground-state phase diagram of this model at half-filling.

\subsection{Exact diagonalization in real space}\label{ED_method}

The topological invariant is one of the most improtant properties to characterize the topological phase transitions, which can be quantified by the Chern number in our model. Given the twisted boundary conditions (TBCs)~\cite{Didier91}, it can be evaluated by~\cite{Niu85},
\begin{align}
  C = \int \frac{d\phi_x d\phi_y}{2 \pi {\rm i}} \left( \langle\partial_{\phi_x}
      \Psi | \partial_{\phi_y} \Psi\rangle - \langle{\partial_{\phi_y}
      \Psi | \partial_{\phi_x} \Psi\rangle} \right),
\label{eq:C}
\end{align}
with $\ket{\Psi}$ being the ground-state wave function. Here $\phi_x$ and $\phi_y$ are the twisted phases along two directions. To avoid the integration of the wave function $\ket{\Psi}$ with respect to the continuous variables, we instead use a discretized version~\cite{Fukui05, Varney11, Zhang13} with intervals $\Delta\phi_x=2\pi/N_x$ and $\Delta\phi_y=2\pi/N_y$. In what follows, ($N_x$, $N_y$)=(20, 20) is adopted to calculate the Chern number.

Other properties used to characterize the critical behavior include the ground-state fidelity metric $g$, which is defined as~\cite{Zanardi06,CamposVenuti07,Zanardi07}
\begin{eqnarray}
g(x,\delta x)\equiv\frac{2}{N }\frac{1-|\langle \Psi(x)|\Psi(x+\delta x)\rangle|}{(\delta x)^2},
\label{eq:g}
\end{eqnarray}
where x represents the parameters V or $\Delta$, and N is the lattice size. $|\Psi(x)\rangle$ [$|\Psi(x+\delta x)\rangle$] is the ground state of $\hat H(x)$ [$\hat H(x+\delta x)$] and we set $\delta x = 10^{-3}$. In addition, the SSO and SCO structure factors can be used to characterize the spin and charge ordered states, respectively, and their definitions in a staggered fashion can be written as
\begin{eqnarray}
S_{\rm SSO} = \frac{1}{N}\sum\limits_{i,j}{(-1)}^{\eta}  \langle (\hat n_{i,\uparrow}-\hat n_{i,\downarrow}) (\hat n_{j,\uparrow}-\hat n_{j,\downarrow})\rangle, \nonumber \\
S_{\rm SCO} = \frac{1}{N}\sum\limits_{i,j}{(-1)}^{\eta}  \langle (\hat n_{i,\uparrow}+\hat n_{i,\downarrow}) (\hat n_{j,\uparrow}+\hat n_{j,\downarrow})\rangle,
\label{eq:S}
\end{eqnarray}
where $\eta = 0$ ($\eta = 1$) if sites $i$ and $j$ are in the same (different) sublattice, i.e., A or B.


\subsection{Mean-field method in momentum space}\label{MF_method}
A variational mean-field method is employed to analyze the ground-state phase diagram of model (\ref{eq:H}), which has been reported in studying the extended Haldane-Hubbard model without lattice potentials~\cite{Shao2021}. For our model, by introducing the operators $a_{\textbf{k},\sigma}^{\dag}=\frac{1}{\sqrt{N}}\sum_{i\in \text{A}}c_{i,\sigma}^{\dag}e^{\mathrm{i}\textbf{k}\cdot\textbf{r}_i}$ and $b_{\textbf{k},\sigma}^{\dag}=\frac{1}{\sqrt{N}}\sum_{i\in \text{B}}c_{i,\sigma}^{\dag}e^{\mathrm{i}\textbf{k}\cdot\textbf{r}_i}$, the Hamiltonian can be expressed as
\begin{align}
\hat H = \hat H_{\text{0}} + \hat H_{\text{I}},
\label{eq:Hk}
\end{align}
where
\begin{align}
\hat H_{\text{0}}=\sum_{\textbf{k},\sigma} \left( m_{+,\sigma}(\textbf{k})a_{\textbf{k},\sigma}^{\dag}a_{\textbf{k},\sigma} + m_{-,\sigma}(\textbf{k})b_{\textbf{k},\sigma}^{\dag}b_{\textbf{k},\sigma}
\notag\right.
\\
\phantom{=\;\;}
\left.-t_1 g(\textbf{k})a_{\textbf{k},\sigma}^{\dag}b_{\textbf{k},\sigma}-t_1 g^*(\textbf{k})b_{\textbf{k},\sigma}^{\dag}a_{\textbf{k},\sigma}\right),
\label{eq:H0}
\end{align}
and
\begin{align}
\hat H_{\text{I}}&=\frac{U}{N}\sum_{\textbf{k},\textbf{k'},\textbf{q}} c_{\textbf{k+q},\uparrow}^{\dag}c_{\textbf{k},\uparrow}c_{\textbf{k}'-\textbf{q},\downarrow}^{\dag}c_{\textbf{k}',\downarrow} \nonumber \\
&+\frac{V}{N}\sum_{\sigma,\sigma'}\sum_{\textbf{k},\textbf{k'},\textbf{q}} g(\textbf{q}) a_{\textbf{k+q},\sigma}^{\dag}a_{\textbf{k},\sigma}b_{\textbf{k}'-\textbf{q},\sigma'}^{\dag}b_{\textbf{k}',\sigma'}.
\label{eq:HI}
\end{align}
Here $g(\textbf{k})=1+e^{-\mathrm{i}\textbf{k}\cdot \textbf{a}_1}+e^{-\mathrm{i}\textbf{k}\cdot \textbf{a}_2}$, and we set
\begin{align}
m_{+,\uparrow}(\mathbf{k})&= +\Delta+m_+(\mathbf{k}) \nonumber \\
m_{+,\downarrow}(\mathbf{k})&= -\Delta+m_+(\mathbf{k})\nonumber \\
m_{-,\uparrow}(\mathbf{k})&= -\Delta+m_-(\mathbf{k}) \nonumber \\
m_{-,\downarrow}(\mathbf{k})&= +\Delta+m_-(\mathbf{k})
\label{eq:Mmp}
\end{align}
with $m_{\pm}(\mathbf{k})= -2t_2[\cos({\bf k}\cdot {\bf a}_1\mp\phi)+\cos({\bf k}\cdot {\bf a}_2\pm\phi)+\cos({\bf k}\cdot ({\bf a}_1-{\bf a}_2)\pm\phi)]$.

By decoupling the four-fermion terms in Eq.(\ref{eq:HI}), the mean-field Hamiltonian can be written as
\begin{equation}
\hat H_{\rm MF}=\hat H_0
+\sum_{\textbf{k}}\psi_{\textbf{k}}^{\dag} \left(
  \begin{array}{cccc}
    \varepsilon_{\uparrow}^{a} & \xi_{\uparrow\uparrow}(\textbf{k}) & \varepsilon_{\uparrow\downarrow}^{a}& \xi_{\uparrow\downarrow}(\textbf{k})\\
    \xi^*_{\uparrow\uparrow}(\textbf{k}) & \varepsilon_{\uparrow}^{b} & \xi^*_{\downarrow\uparrow}(\textbf{k}) &\varepsilon_{\uparrow\downarrow}^{b}\\
    (\varepsilon_{\uparrow\downarrow}^{a})^* & \xi_{\downarrow\uparrow}(\textbf{k}) & \varepsilon_{\downarrow}^{a} & \xi_{\downarrow\downarrow}(\textbf{k})\\
    \xi^*_{\uparrow\downarrow}(\textbf{k}) & (\varepsilon_{\uparrow\downarrow}^{b})^* & \xi^*_{\downarrow\downarrow}(\textbf{k}) & \varepsilon_{\downarrow}^{b}\nonumber
  \end{array}
\right)\psi_{\textbf{k}},
\end{equation}
where  $\psi^\dagger_{\textbf{k}}=[a^\dagger_{\textbf{k},\uparrow},b^\dagger_{\textbf{k},\uparrow},a^\dagger_{\textbf{k},\downarrow},b^\dagger_{\textbf{k},\downarrow}]$ represents the basis for each lattice momentum ${\bf k}$ and
\begin{align}
\xi_{\sigma\sigma'}(\textbf{k})&=-\frac{V}{N}\sum_{\textbf{q}}g(\textbf{k}-\textbf{q})\langle b_{\textbf{q},\sigma'}^{\dag} a_{\textbf{q},\sigma}\rangle_{\rm MF},\nonumber \\
\varepsilon_{\sigma}^{a}&=Un_{-\sigma}^{a}+3V\sum_{\sigma'} n_{\sigma'}^{b},\nonumber \\
\varepsilon_{\sigma}^{b}&=Un_{-\sigma}^{b}+3V\sum_{\sigma'} n_{\sigma'}^{a},\nonumber \\
\varepsilon_{\uparrow\downarrow}^{a}&=-\frac{U}{N}\sum_{\textbf{q}}\langle a_{\textbf{q},\downarrow}^{\dag} a_{\textbf{q},\uparrow}\rangle_{\rm MF},\nonumber \\
\varepsilon_{\uparrow\downarrow}^{b}&=-\frac{U}{N}\sum_{\textbf{q}}\langle b_{\textbf{q},\downarrow}^{\dag} b_{\textbf{q},\uparrow}\rangle_{\rm MF},
\end{align}
with densities $n_{\sigma}^{a}=\frac{1}{N}\sum_{\textbf{q}}\langle a_{\textbf{q},\sigma}^{\dag} a_{\textbf{q},\sigma}\rangle_{\rm MF}$ and $n_{\sigma}^{b}=\frac{1}{N}\sum_{\textbf{q}}\langle b_{\textbf{q},\sigma}^{\dag} b_{\textbf{q},\sigma}\rangle_{\rm MF}$.
The above mean-field equations can be solved self-consistently by making use of the variational mean-field approach. Once the free energy has converged, the SSO and SCO order parameters can be obtained by
\begin{eqnarray}
\mathcal{O}_{\rm SSO} &=& \left|\frac{1}{2}\left(\langle \vec  S_A \rangle_{\rm MF} - \langle \vec S_B\rangle_{\rm MF} \right)\right|,\nonumber \\
\mathcal{O}_{\rm SCO} &=& \left|\left( n^{A}_{\uparrow}+ n^{A}_{\downarrow})-( n^{B}_{\uparrow}+ n^{B}_{\downarrow}\right)\right|.
\label{eq:O}
\end{eqnarray}
Here $\vec S_i = \frac{1}{2}\sum_{\alpha\beta}c_{i\alpha}^\dagger \vec\sigma_{\alpha\beta}c^{\phantom{\dagger}}_{i\beta}$, and $\vec \sigma = (\sigma^x, \sigma^y, \sigma^z)$ is the vector of spin-1/2 Pauli matrices. Meanwhile, we use the discrete formulation in its multiband (non-Abelian) version to compute the Chern number~\cite{Fukui05}.

Besides, similar to the definitions in Ref.~\cite{He_2024, silva2023}, we define the effective potential differences for spin-$\uparrow$ and spin-$\downarrow$ electrons in the MF method as
\begin{eqnarray}
\Delta_{\rm MF}^{\uparrow} &=& \frac{|\varepsilon_{\uparrow}^{a}-\varepsilon_{\uparrow}^{b}+2\Delta|}{2},\nonumber \\
\Delta_{\rm MF}^{\downarrow} &=& \frac{|\varepsilon_{\downarrow}^{a}-\varepsilon_{\downarrow}^{b}-2\Delta|}{2}.
\label{potential}
\end{eqnarray}
Notice that $+2\Delta$ in $\Delta_{\rm MF}^{\uparrow}$ and $-2\Delta$ in $\Delta_{\rm MF}^{\downarrow}$ come from the opposite potential differences for spin $\uparrow$ and $\downarrow$ species.


\section{Results and analysis}\label{sec:results}

\subsection{Results of the exact diagonalization method}\label{sec:ED}
\begin{figure}[t]
\centering
\includegraphics[width=0.48\textwidth]{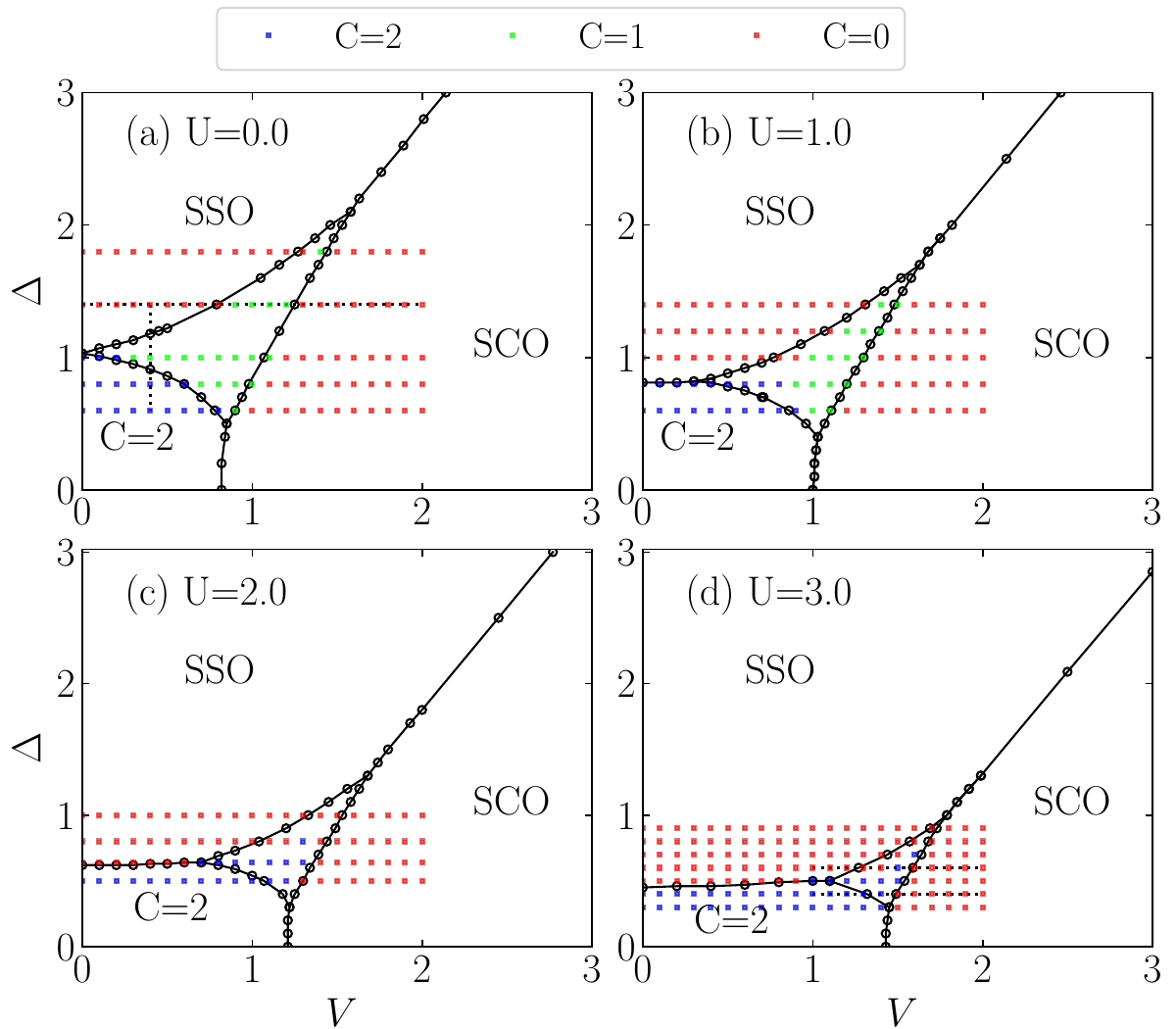}
\caption{Phase diagram in the parametric space ($V$, $\Delta$) of the model (\ref{eq:H}) based on the results of fidelity metric $g$ with (a) $U=0.0$, (b) $U=1.0$, (c) $U=2.0$ and (d) $U=3.0$. The blue, green and red squares indicate results of Chern number $C=2$, $C=1$ and $C=0$, respectively. The black dashed lines in (a) and (d) denote the parameters we choose to show more details below.
}
\label{fig_1}
\end{figure}

In the ED calculations, we choose the $12$A cluster whose reciprocal lattice encompasses the $\Gamma$, K, K$'$ points, and one pair of M points, see more details in Ref.~\cite{Shao2021}. It has been shown that including the high-symmetry K and K$'$ points is crucial for studying the quantum phase transition in interacting Haldane models~\cite{Shao2021, Varney10, Varney11}. The periodic boundary condition is utilized for the calculation of all properties and when determining the Chern numbers, we make use of translational symmetries to reduce the size of the Hilbert space. The ($V$, $\Delta$) phase diagrams are shown in Figs.~\ref{fig_1}(a), \ref{fig_1}(b), \ref{fig_1}(c) and \ref{fig_1}(d) with $U=0.0$, $U=1.0$, $U=2.0$ and $U=3.0$, respectively. In Fig.~\ref{fig_1}, the phase transition points (black circles) are identified based on the positions of the peaks of fidelity metric $g$.

Let us first focus on the $U=0.0$ case in Fig.~\ref{fig_1}(a), where the Chern number results match the phase diagram obtained from the fidelity metric very well. Here we use the blue, green and red squares to represent the Chern number $C=2$, $C=1$ and $C=0$, respectively. We can observe from Fig.~\ref{fig_1} (a) that the Chern insulator phase with $C=2$ is present in a region with small $V$ and $\Delta$, while the SSO (SCO) phase with $C=0$ dominates the system when $\Delta$ ($V$) is sufficiently large. Interestingly, an intermediate phase with $C=1$ can be observed surrounded by the three expected phases. This is different from the phase diagram in Ref.~\cite{Vanhala2016}, where the $C=1$ phase is sandwiched between a band insulator and Mott insulator. Notice that their charge ordered phase (band insulator) is governed by the ionic potential $\Delta_{\rm AB}$ which is not spin-dependent, and their spin ordered phase (Mott insulator) is governed by the on-site Hubbard interaction $U$. In contrast, in our case, the staggered charge order (SCO) phase is governed by the nearest-neighbor interaction $V$ and the staggered spin order (SCO) phase is governed by the spin-dependent lattice potential $\Delta_{\sigma,\alpha}$. We propose that the different competition mechanisms result in the different locations of the $C=1$ phase.


The impact of the on-site interaction $U$ on the phase diagram is depicted in Figs.~\ref{fig_1}(b), \ref{fig_1}(c) and \ref{fig_1}(d). In Fig.~\ref{fig_1}(b) with $U=1.0$, there is a discrepancy between the results of the Chern number and the phase diagram obtained from the fidelity metric. Moreover, results of $C=1$ (green points) can not even be observed in Figs.~\ref{fig_1}(c) and \ref{fig_1}(d) with $U=2.0$ and $U=3.0$, respectively. On the other hand, the area of the ``intermediate phase'' identified by the fidelity metric decreases but does not vanishes as $U$ increases from $0.0$ to $3.0$. We discuss the ($U$, $\Delta$) phase diagram with $V=0$ in Appendix \ref{Appendix_U} to better understand the influence of $U$ on the ($V$, $\Delta$) phase diagrams. We must have a statement here that in our ED discussions, the $C=1$ phase is equivalent to the ``intermediate phase'' only when $U=0.0$, and they are not equivalent when $U$ is finite because of the discrepancies. Similar discussions have also been reported in Refs.~\cite{Shao2021, shao23} and they found that it is better to use the fidelity results to characterize the phase transitions in the interacting Haldane models because the discretized method to calculate Chern number suffers from the finite-size effect more severely. The MF results in Sec.~\ref{sec:MF} also support the existence of $C=1$ in the case of $U=3.0$. Even so, we still suggest that the question of whether the $C = 1$ phase can exist with $U$ increased to $3.0$ may be open. This issue is worth further investigation, especially considering that in materials $U$ is generally larger than $V$.

\begin{figure}[t]
\centering
\includegraphics[width=0.48\textwidth]{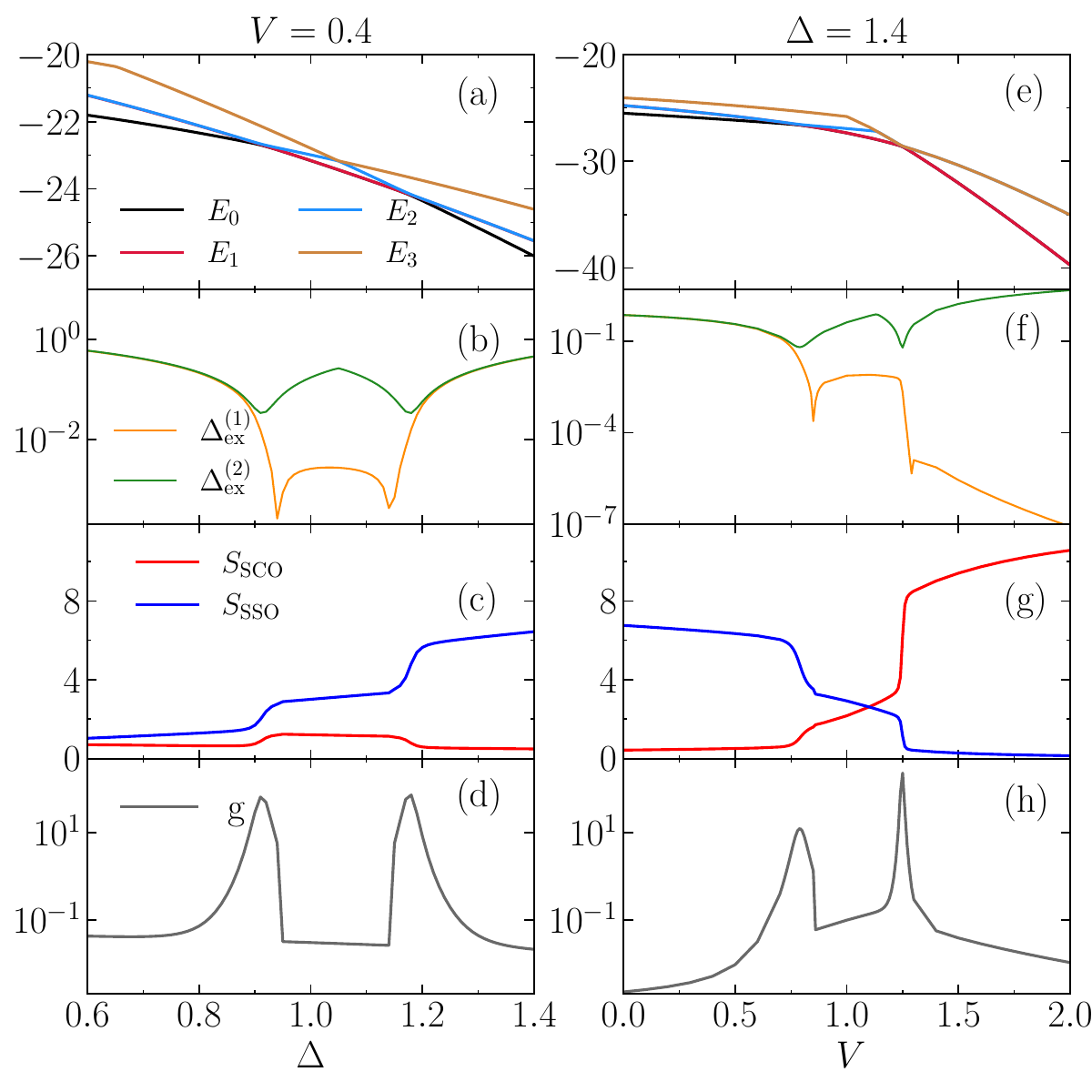}
\caption{(a)(e) Four lowest-lying energy levels $E_{\alpha}$, (b)(f) the excitation gaps $\Delta_{\text{ex}}^{(\alpha)}$, (c)(g) the structure factors $S_{\text{SSO/SCO}}$, and (d)(h) the fidelity metric $g$ of the model (\ref{eq:H}) with $V=0.4$ on the left panels and $\Delta=1.4$ on the right panels. The on-site interaction $U=0$ and the parameters are corresponding to the black dashed lines in Fig.~\ref{fig_1}(a).
}
\label{fig_2}
\end{figure}

To elucidate the critical behaviors of the phase diagram in Fig.~\ref{fig_1}(a) with $U=0.0$, we focus on the two black dashed lines with $V=0.4$ and $\Delta=1.4$ to calculate the relevant properties. That is, the left and right panels in Fig.~\ref{fig_2} are corresponding to the results of $V=0.4$ and $\Delta=1.4$, respectively. The first four lowest-lying energy levels $E_{\alpha}$ ($\alpha=0,1,2,3$ and $\alpha=0$ represents the ground state) are obtained by employing the Arnoldi~\cite{Lehoucq97arpack} method, see Figs.~\ref{fig_2}(a) and \ref{fig_2}(e). In Fig.~\ref{fig_2}(a), the system traverses the $C=2$, $C=1$ and SSO phases as $\Delta$ increases from $0.6$ to $1.4$, and two level crossings between the ground state and one excited state are observed at $\Delta\approx0.9$ and $\Delta\approx1.2$. In Fig.~\ref{fig_2}(e), the system crosses the SSO, $C=1$ and SCO phases as $V$ increases from $0.0$ to $2.0$, and two level crossings occur at $V\approx0.75$ and $V\approx1.25$. To provide a more intuitive representation, we exhibit the excitation gaps $\Delta_{\rm ex}^{(1)}=E_1-E_0$ and $\Delta_{\rm ex}^{(2)}=E_2-E_0$, as shown in Figs.~\ref{fig_2}(b) and \ref{fig_2}(f). It is evident that the excitation gaps exhibit local minimum values at the critical points where level crossings occur. It is worth noting that in Fig.~\ref{fig_2}(f), $\Delta_{\rm ex}^{(1)}$ becomes very small when $V>1.25$, owing to the nearly degenerate ground state and first excited state in SCO phase. Another feature to characterize the phase transitions is the change of the SCO or SSO structure factor ($S_{\text{SCO}}$ or $S_{\text{SSO}}$), see Figs.~\ref{fig_2}(c) and \ref{fig_2}(g). Values of $S_{\text{SCO}}$ ($S_{\text{SSO}}$) in $C=1$ phase is larger than those in $C=2$ phase but smaller than those in the SCO (SSO) phase, indicating again that $C=1$ is an intermediate state resulting from the interplay between topology, electronic correlations and lattice potentials. Finally, sharp peaks of the ground-state fidelity metrics $g$ in Figs.~\ref{fig_2}(d) and \ref{fig_2}(h) can be observed and used to characterize the critical points. Besides, for $U=3.0$ in Fig.~\ref{fig_1}(d), similar calculations and discussions along the black dashed lines with $\Delta=0.4$ and $\Delta=0.6$ are detailed in Appendix~\ref{Appendix_U3}.

Before closing this section, we would like to note that the finite-size effect in ED method is discussed in Appendix~\ref{Appendix_6site}.

\subsection{Results of the mean-field method}\label{sec:MF}

\begin{figure}[t]
\centering
\includegraphics[width=0.45\textwidth]{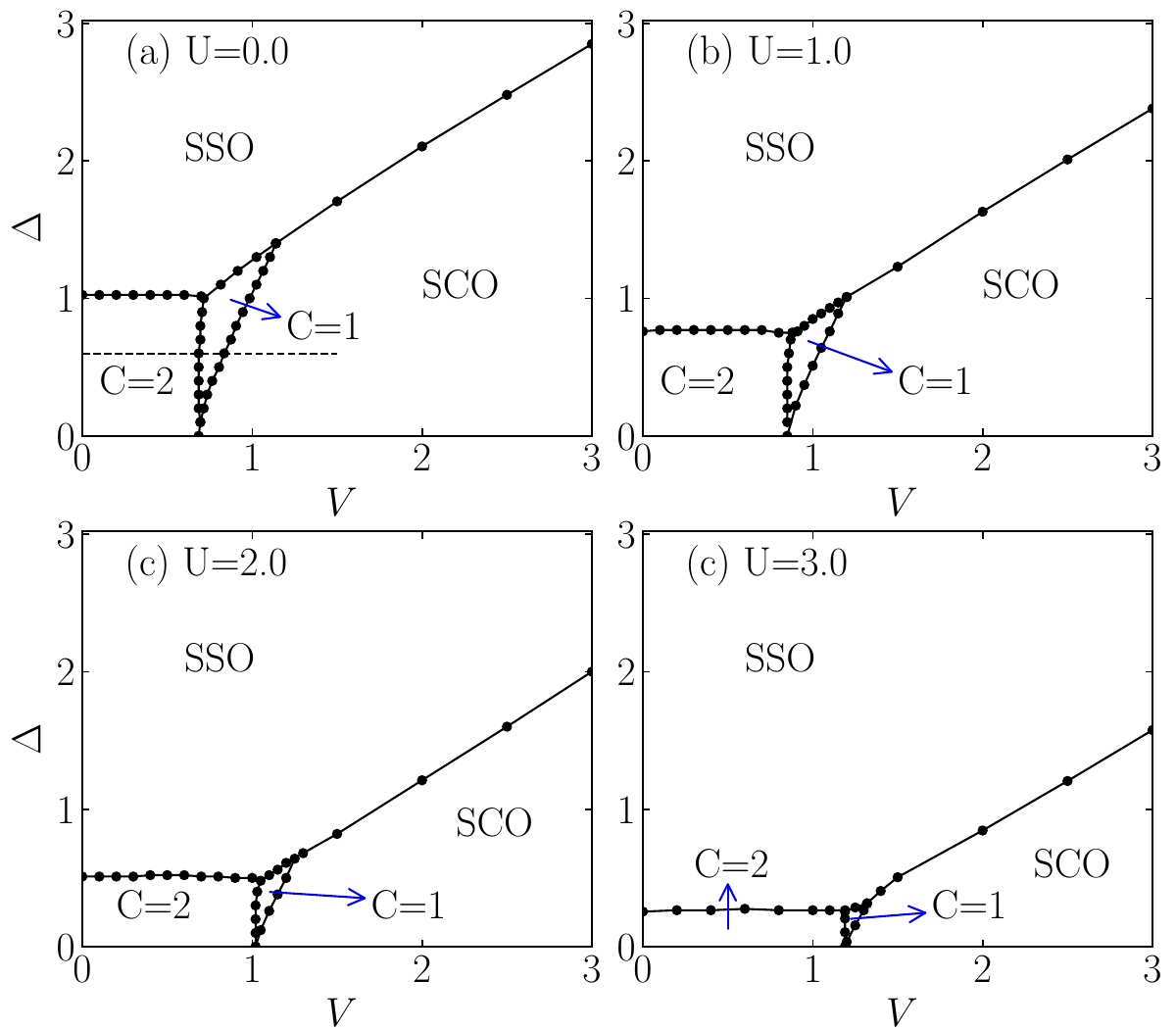}
\caption{Mean-field phase diagram in the parametric space ($V$, $\Delta$) of the model (\ref{eq:H}) based on the results of Chern number $C$ and the order parameters ($\mathcal{O}_{\text{SCO}}$ and $\mathcal{O}_{\text{SSO}}$), with (a) $U=0.0$, (b) $U=1.0$, (c) $U=2.0$ and (d) $U=3.0$.
}
\label{fig_3}
\end{figure}

\begin{figure}[t]
\centering
\includegraphics[width=0.45\textwidth]{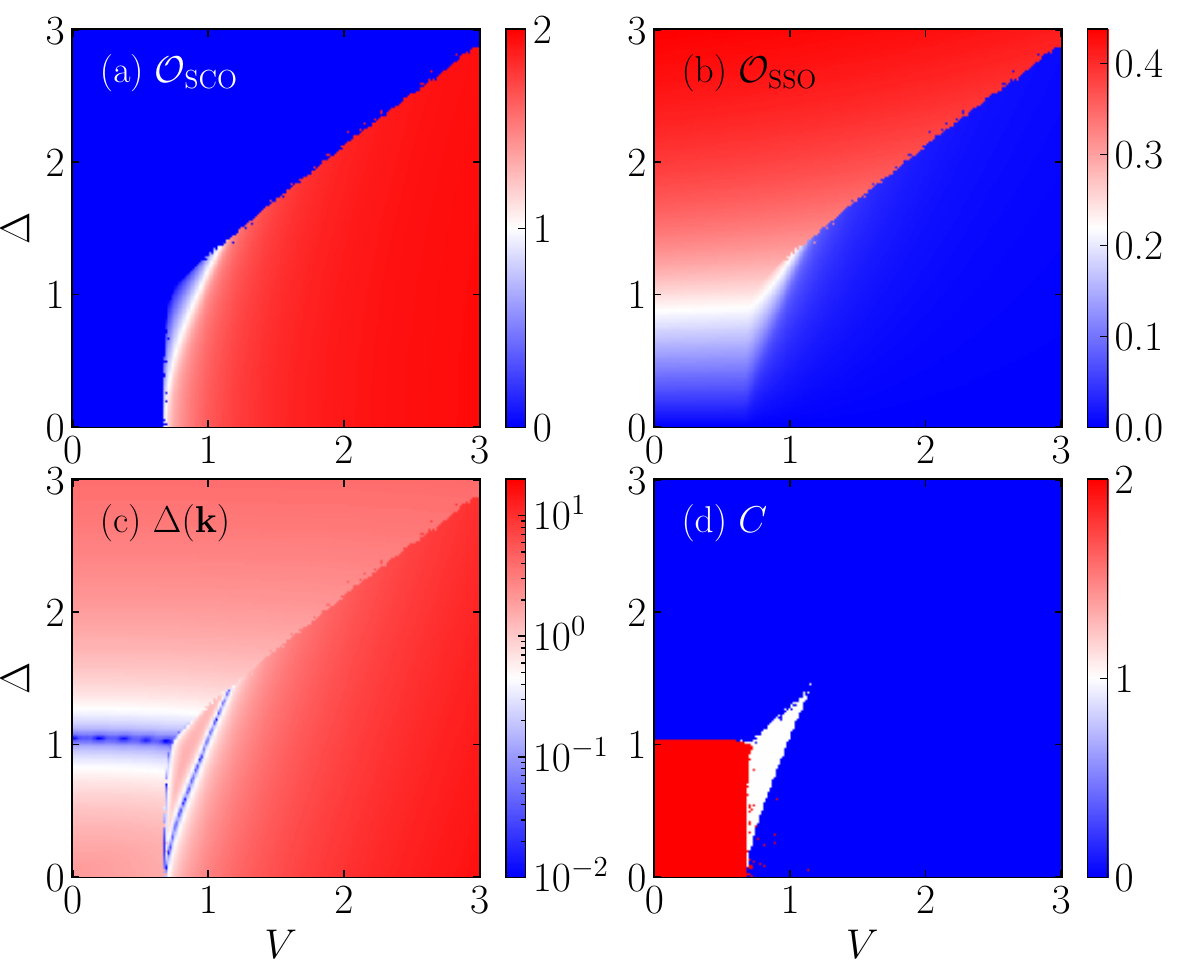}
\caption{Contour plots of (a) the SCO order parameter, (b) the SSO order parameter, (c) the band gap and (d) the Chern number, as a function of $V$ and $\Delta$. The mean-field approach is adopted for the calculations of the model (\ref{eq:H}) with $U=0.0$.
}
\label{fig_4}
\end{figure}

To contrast the exact results on small lattice above, we now report the outcomes of the mean-field method. We use the $180\times180$ lattice to calculate the band gap and the order parameters ($\mathcal{O}_{\text{SCO}}$ and $\mathcal{O}_{\text{SSO}}$), while a $20\times20$ lattice is adopted for the calculations of Chern number. Despite the differences in lattice size, the critical points identified by the Chern number and other properties are consistent, as will be shown in Fig.~\ref{fig_4}. Phase diagrams with regard to the parameters $V$ and $\Delta$ are shown in Figs.~\ref{fig_3}(a), \ref{fig_3}(b), \ref{fig_3}(c) and \ref{fig_3}(d) with $U=0.0$, $U=1.0$, $U=2.0$ and $U=3.0$, respectively. In comparison to the phase diagrams of ED in Fig.~\ref{fig_1}, similar features can be observed: SSO and SCO dominate the system for large $\Delta$ and $V$, respectively, leaving the $C=2$ phase in small ($V$, $\Delta$) region; the $C=1$ phase are surrounded by the above three phases. While differences between the MF and ED phase diagrams include: \\
(1) In the MF method, the increased interaction $U$ does not eliminate the $C=1$ phase at least for $U\leq3.0$.\\
(2) The $C=1$ phase in Fig.~\ref{fig_1}(a) can exist with very small $V$, while in small $\Delta$ region ($\Delta\leq0.5$) it vanishes. This is opposite to the MF results in Fig.~\ref{fig_3}(a), where the $C=1$ phase can exist in small $\Delta$ region rather than small $V$ region.

Taking the $U=0.0$ case in Fig.~\ref{fig_3}(a) as an example, we show in Figs.~\ref{fig_4}(a), \ref{fig_4}(b), \ref{fig_4}(c) and \ref{fig_4}(d) the results of the SCO order parameter ($\mathcal{O}_{\text{SCO}}$), the SSO order parameter ($\mathcal{O}_{\text{SSO}}$), the band gap ($\Delta(\textbf{k})$) and the Chern number ($C$), respectively, as a function of $V$ and $\Delta$. It is noteworthy that the phase boundary between SSO and SCO can only be identified from the order parameters, and other phase boundaries are characterized by the results of Chern number, as seen in Fig.~\ref{fig_4}(d). Despite some defect points in Fig.~\ref{fig_4}(d), possibly attributed to the severe quantum fluctuation near the critical points, the phase boundaries can be readily identified. In Fig.~\ref{fig_4}(a), we observe that $\mathcal{O}_{\text{SCO}}$ vanishes in the SSO and $C=2$ phases, while the $C=1$ phase can be regarded as a zone of transition with $\mathcal{O}_{\text{SCO}}$ changing from $0.0$ to $2.0$. For $\mathcal{O}_{\text{SSO}}$ in Fig.~\ref{fig_4}(b), the $C=1$ phase remains a transition region with finite values but smaller that those in SSO phase. These features are similar to the structure factors of ED results shown in Figs.~\ref{fig_2}(c) and \ref{fig_2}(g). To contrast the excitation gaps in ED results, we show the band gap defined as $\Delta(\textbf{k})$=min [$E_{2}(\textbf{k}^{\prime})-E_{1}(\textbf{k}^{\prime})$] in Fig.~\ref{fig_4}(c). The rapid drop or closing of the gap size can be observed at the critical values, except for the phase boundary between SSO and SCO, where only discontinuity in gap size appears. This can be connected with the general picture that the change of a topological invariant is always accompanied by a single-particle gap closing. While for the phase transition from SSO to SCO side, the Chern number does not change ($C=0$). This may explain the discontinuity of gap size instead of gap closing, considering that the SSO and SCO states are all gapped phases with different gap sizes.

\begin{figure}[t]
\centering
\includegraphics[width=0.45\textwidth]{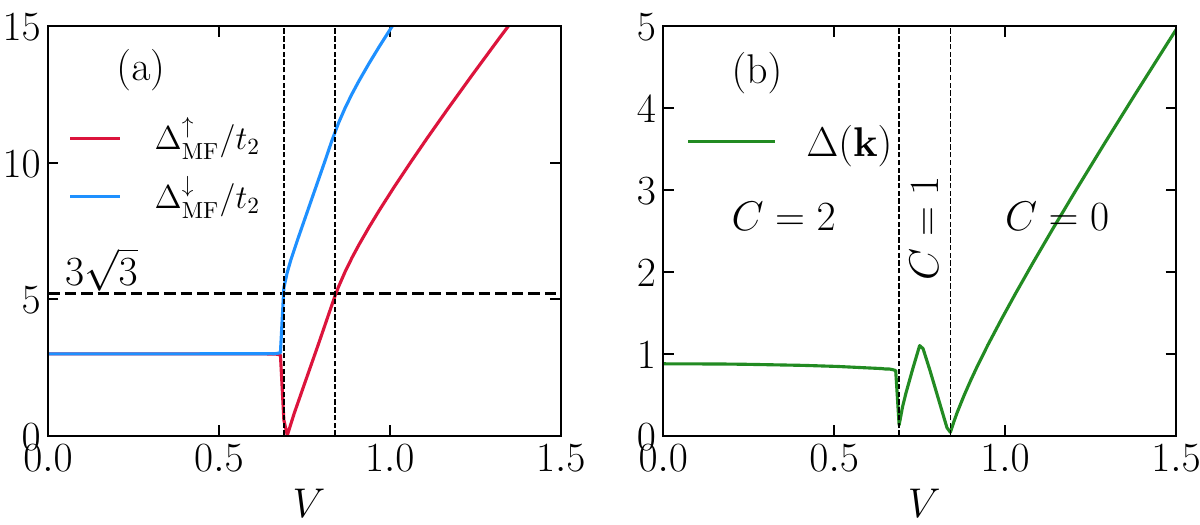}
\caption{(a) The effective potential differences in units of $t_2$ $\Delta_{\rm MF}^{\uparrow}/t_2$ and $\Delta_{\rm MF}^{\downarrow}/t_2$, and (b) the band gap as a function of $V\in[0.0, 1.5]$ (see the black dashed line in Fig.~\ref{fig_3}(a)). Other parameters: $U=0$ and $\Delta=0.6$. The horizontal dashed line in (a) denotes the value of $3\sqrt3$ in y axis, and the vertical dashed lines in (a) and (b) separate the $C=2$, $C=1$ and $C=0$ regions.
}
\label{fig_5}
\end{figure}

To analyze the origin of the $C=1$ phase, we define the effective potential differences $\Delta_{\rm MF}^{\uparrow}$ and $\Delta_{\rm MF}^{\downarrow}$ in Eq.~(\ref{potential}), and here we show $\Delta_{\rm MF}^{\uparrow}/t_2$ and $\Delta_{\rm MF}^{\downarrow}/t_2$ as a function of $V$ in Fig.~\ref{fig_5}(a). We set $U=0$, $\Delta=0.6$ and $V\in[0.0, 1.5]$, see the black dashed line in Fig.~\ref{fig_3}(a). It is known that in the noninteracting spinless Haldane model, the Chern number $C$ equals $1$ under the condition $|\Delta|/t_2<3\sqrt3$, and $C$ equals $0$ when $|\Delta|/t_2>3\sqrt3$. From Fig.~\ref{fig_5}(a) we can observe that:

$\Delta_{\rm MF}^{\uparrow}/t_2<3\sqrt3$ and $\Delta_{\rm MF}^{\downarrow}/t_2<3\sqrt3$ when $C=2$;\\

$\Delta_{\rm MF}^{\uparrow}/t_2>3\sqrt3$ and $\Delta_{\rm MF}^{\downarrow}/t_2<3\sqrt3$ when $C=1$;\\

$\Delta_{\rm MF}^{\uparrow}/t_2>3\sqrt3$ and $\Delta_{\rm MF}^{\downarrow}/t_2>3\sqrt3$ when $C=0$.\\
Here the vertical dashed lines in Fig.~\ref{fig_5}(a) separate the $C=2$, $C=1$ and $C=0$ regions. In Fig.~\ref{fig_5}(b), we show the band gap $\Delta(\textbf{k})$ as a function of $V$, where two gap closures can be found at the critical points, denoted by the vertical dashed lines identical to those in Fig.~\ref{fig_5}(a). These features indicates that the formation of the $C = 1$ phase is due to the spontaneous SU(2) symmetry breaking, with only one spin species in topological state.

\section{Summary and Discussion}\label{sec:conclusion}

In summary, we investigated the interacting spinful Haldane model at half-filling on the honeycomb lattice with spin-dependent sublattice potentials. By employing the exact-diagonalization (ED) and mean-field (MF) methods, we obtained similar ground-state phase diagrams for the nearest-neighbor interaction $V\in[0,3]$ and sublattice potential difference $\Delta\in[0,3]$. The staggered spin order (SSO) and staggered charge order (SCO) phases dominate in the large $\Delta$ and $V$ region, respectively; the $C=2$ phase prevails when both $V$ and $\Delta$ are small enough; and an intermediate phase with $C=1$ is surrounded by the aforementioned three phases. Except for the change in the topological invariant, other features such as the closure of the excitation gap, changes in structure factors, and peaks in the fidelity metric were also observed in the ED results. Meanwhile, we examined the mean-field Chern number, band gap and local order parameters, also providing clear evidence for the existence of the $C=1$ phase. Especially, by analyzing the effective potential differences $\Delta_{\rm MF}^{\uparrow}$ and $\Delta_{\rm MF}^{\downarrow}$, the origin of this exotic phase is also attributed to the spontaneous SU(2) symmetry breaking.

However, the interaction $U$ suppresses the $C=1$ phase, raising an open question about its potential existence as $U$ increased to $3.0$. In the ED approach, $C=1$ can not be found when $U=3.0$ even though the ``intermediate phase'' can still be identified from the fidelity metric results. In the MF approach, the $C=1$ phase preserves for $U=3.0$, albeit with a very small area.

Finally, we would like to note that in previous investigations, the staggered spin order (SSO) has been denoted as the antiferromagnetic (AF)\cite{Jiang18, Wang2019} or spin-density-wave (SDW) state~\cite{He_2011, He_2024, silva2023, shao23, Shao2021}, and the staggered charge order (SCO) has been labeled as the band insulator (BI)\cite{Wang2019,Vanhala2016} or charge-density-wave (CDW)\cite{Wang2019, He_2024, silva2023, shao23, Shao2021} state. While in this paper, we refer to them as SSO and SCO to stress their staggered characteristics on A and B sublattices.

\begin{acknowledgments}
The authors acknowledge insightful discussions with E. V. Castro, R. Mondaini, H. Lu and S. Hu.

C. S. acknowledges support from the National Natural Science Foundation of China (NSFC; Grant No. 12104229) and the Fundamental Research Funds for the Central Universities (Grant No. 30922010803).
H.-G. L. acknowledges support from NSFC (Grants No. 11834005 and No. 12247101), and the National Key Research and Development Program of China (Grant No. 2022YFA1402704).
\end{acknowledgments}

\appendix

\section{the ED phase diagram in the parametric space ($U$, $\Delta$) with $V=0$} \label{Appendix_U}

\begin{figure}[t]
\centering
\includegraphics[width=0.48\textwidth]{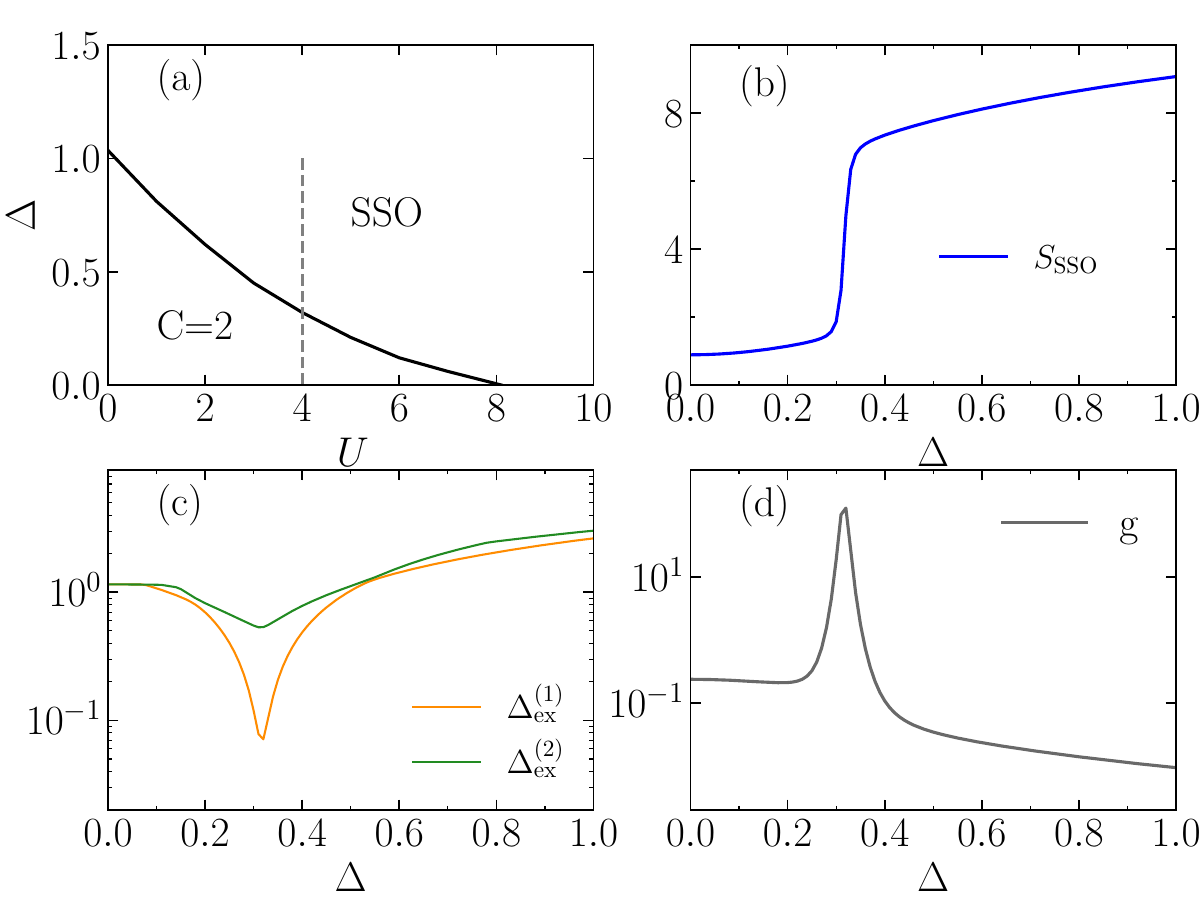}
\caption{(a) The exact-diagonalization (ED) phase diagram of the model (\ref{eq:H}) in the parametric space($U$, $\Delta$) with $V=0$. The black dashed line denotes the parameters we choose to show the structure factor $S_{\text{SSO}}$ in (b), the excitation gaps $\Delta_{\text{ex}}^{(\alpha)}$ in (c), and the fidelity metric $g$ in (d).
}
\label{fig_A1}
\end{figure}

To better understand the impact of the interaction parameter $U$ on the phase diagram, we present the ED phase diagram of our model as a function of $U$ and $\Delta$ in Fig.~\ref{fig_A1}(a), with $V=0$. It can be observed that both $U$ and $\Delta$ favor the SSO order, and $\Delta$ is more efficient in suppressing the $C=2$ phase compared to $U$. Specifically, for $\Delta=0$, the critical point is located at $U=8.1$, while for $U=0$, the critical point is located at $\Delta=1.03$. This explains why increasing $U$ from $0$ to $3$ does not eliminate the intermediate phase in the ($V$, $\Delta$) phase diagram, as depicted in Fig.~\ref{fig_1} of the main text.

Furthermore, we choose $U=4$ [black dashed line in Fig.~\ref{fig_A1}(a)] to plot the structure factor of SSO in Fig.~\ref{fig_A1}(b), the excitation gaps in Fig.~\ref{fig_A1}(c) and the fidelity metric in Fig.~\ref{fig_A1}(d). The rapid change of $S_{\text{SSO}}$, the closure of $\Delta_{\text{ex}}^{(1)}$ and the sharp peak of the fidelity metric $g$ can be observed, respectively. Notably, no intermediate phase can be identified from these properties, indicating that the $C=1$ phase results from the competition between $V$ and $\Delta$.

\section{details of some properties in ED method with $U=3$} \label{Appendix_U3}

\begin{figure}[t]
\centering
\includegraphics[width=0.48\textwidth]{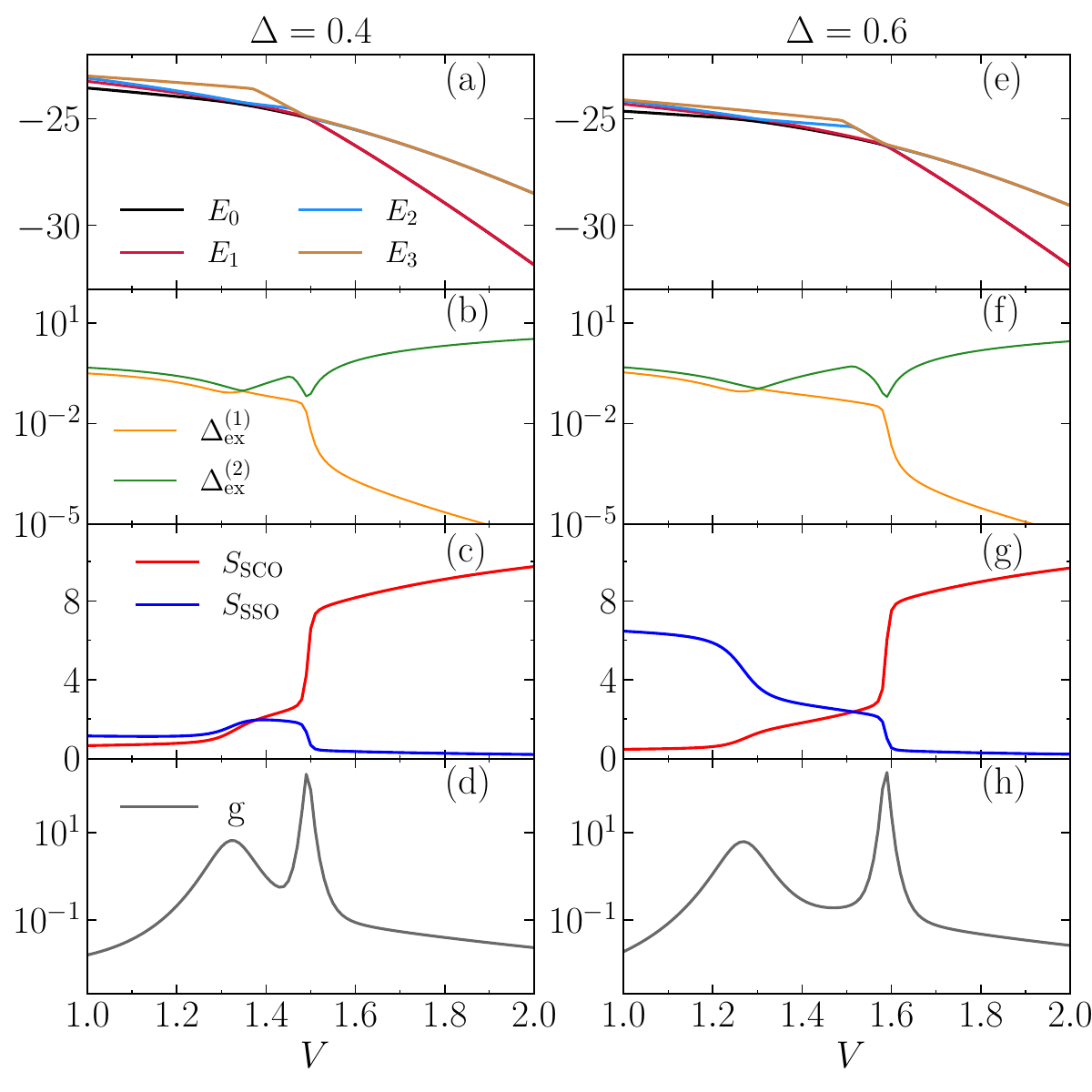}
\caption{(a)(e) Four lowest-lying energy levels $E_{\alpha}$, (b)(f) the excitation gaps $\Delta_{\text{ex}}^{(\alpha)}$, (c)(g) the structure factors $S_{\text{SSO/SCO}}$, and (d)(h) the fidelity metric $g$ of the model (\ref{eq:H}) with $\Delta=0.4$ on the left panels and $\Delta=0.6$ on the right panels. The on-site interaction $U=3$ and the parameters are corresponding to the black dashed lines in Fig.~\ref{fig_1}(d).
}
\label{fig_A2}
\end{figure}

In Fig.~\ref{fig_A2}, considering the case of $U=3.0$, we choose $\Delta=0.4$ (left panel) and $\Delta=0.6$ (right panel) to illustrate the relevant properties. The line of $\Delta=0.4$ ($\Delta=0.6$) in Fig.~\ref{fig_1}(d) crosses the $C=2$ phase (SSO phase), the ``intermediate phase'' and the SCO phase as $V$ increases from $1.0$ to $2.0$ [see black dashed lines in Fig.~\ref{fig_1}(d)]. Examining the fidelity metrics $g$ in Figs.~\ref{fig_A2}(d) and \ref{fig_A2}(h), we observe sharp peaks between the ``intermediate phase'' and the SCO phase, located at $V\approx1.5$ and $V\approx1.6$, respectively. In the vicinity of the locations of the shape peaks, level crossings between the ground state and the second excited state can be observed [Figs.~\ref{fig_A2}(a) and \ref{fig_A2}(e)], as well as the minimum values of the second excitation gap $\Delta_{\text{ex}}^{(2)}$ [Figs.~\ref{fig_A2}(b) and \ref{fig_A2}(f)] and sudden changes in structure factors [Figs.~\ref{fig_A2}(c) and \ref{fig_A2}(g)].  Notice that in SCO phase the ground state and first excited state are nearly degenerate.

On the other hand, only ``hump''s are observed at the phase boundaries between the $C=2$ (or SSO) and the ``intermediate phase'', locating at $V\approx1.32$ in Fig.~\ref{fig_A2}(d) [$V\approx1.27$ in Fig.~\ref{fig_A2}(h)]. Near the positions of the ``hump''s, level crossings between the ground state and the first excited state [Figs.~\ref{fig_A2}(a) and \ref{fig_A2}(e)], the local minimum values of the first excitation gap $\Delta_{\text{ex}}^{(1)}$ [Figs.~\ref{fig_A2}(b) and \ref{fig_A2}(f)] and smooth changes in structure factors [Figs.~\ref{fig_A2}(c) and \ref{fig_A2}(g)] can be observed. Similar features have been discussed in Refs.~\cite{Shao2021, shao23}, where the reason is attributed to the finite-size effect, and it can be mitigated by employing different clusters or twisted boundary conditions.

\section{the ED phase diagram in the parametric space ($V$, $\Delta$) on a 6-site lattice} \label{Appendix_6site}

\begin{figure}[t]
\centering
\includegraphics[width=0.48\textwidth]{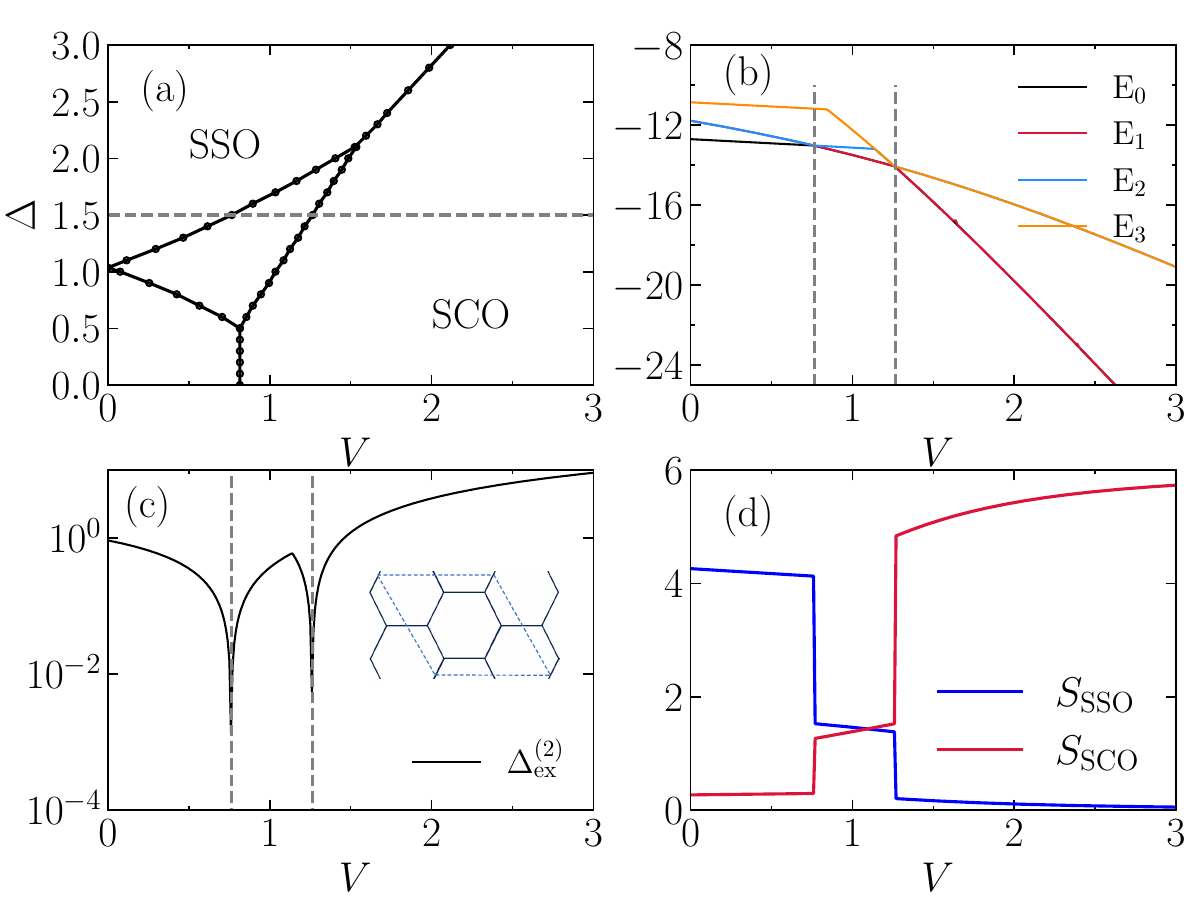}
\caption{(a) The exact-diagonalization (ED) phase diagram of the model (\ref{eq:H}) in the parametric space($V$, $\Delta$) with $U=0$, on a $6$-site cluster shown in (c). The black dashed line in (a) denotes the parameters we choose to show the spectrum in (b), the excitation gap $\Delta_{\text{ex}}^{(2)}$ in (c), and the structure factors $S_{\rm SSO/SCO}$ in (d).}
\label{fig_A3}
\end{figure}

In the beginning of Sec.~\ref{sec:results}, we mentioned that we utilized a $12$A cluster, whose reciprocal lattice contains the K points~\cite{Shao2021}, to study the phase transitions of the model~(\ref{eq:H}) in the main text. Now we would like to discuss the finite-size effect in our ED study. Is is known that during the phase transition of noninteracting Haldane model, the gap closes at the K points. It's important to note that the condition for reciprocal lattice of a cluster to contain the K points can only be satisfied when the site number of the cluster is a multiple of $6$\cite{Varney10}. This condition has been demonstrated to be crucial for characterizing the topological phase transitions in the interacting Haldane models\cite{Varney10,Varney11,Shao2021,shao23}. Unfortunately, due to limitations in our current computational resources, we are unable to access an $18$A cluster. Therefore, we conducted our phase diagram analysis on the $6$A cluster instead, as shown in subplot of Fig.~\ref{fig_A3}(c).

As expected, an intermediate phase can be observed in the ($V$, $\Delta$) phase diagram in Fig.~\ref{fig_A3}(a) and we select the dashed line with $\Delta=1.5$ to illustrate more detailed properties. In Fig.~\ref{fig_A3}(b), we depict four lowest-lying energy levels, and the level crossings at two critical points can be observed. It's worth noting that the ground state $E_0$ and first excited state $E_1$ are degenerate in the SCO and intermediate phase, while the first excited state $E_1$ and the second excited state $E_2$ are degenerate in the SSO phase. As a result, it is more appropriate to use the second excitation gap, $\Delta_{\rm ex}^{(2)}$, to characterize the gap closure at phase transition points, as shown in Fig.~\ref{fig_A3}(c). Additionally, we present the structure factors $S_{\rm SSO}$ and $S_{\rm SCO}$ in Fig.~\ref{fig_A3}(d), where the transition nature of the intermediate phase is clearly visible.


%

\end{document}